\documentclass[twocolumn,showpacs,preprintnumbers,amsmath,amssymb]{revtex4}
\usepackage{graphicx}
\usepackage{dcolumn}
\usepackage{bm}
\begin{document}
\title{Analysis of shot noise suppression in mesoscopic cavities in a magnetic
field}
\author{P.~Marconcini}
\author{M.~Macucci}
\author{G.~Iannaccone}
\author{B.~Pellegrini}
\author{G.~Marola}
\affiliation{Dipartimento di Ingegneria dell'Informazione,
Universit\`a di Pisa\\
Via Caruso, I-56122 Pisa, Italy}
\date{\today}
\pacs{72.70.+m, 73.23.-b, 75.75.+a}
\begin{abstract}
We present a numerical investigation of shot noise suppression in mesoscopic
cavities and an intuitive
semiclassical explanation of the behavior observed in the presence of an 
orthogonal magnetic field. In particular, we  
conclude that the decrease of shot noise for increasing magnetic field 
is the result of the interplay between the diameter of classical cyclotron
orbits and the width of the apertures defining the cavity. Good agreement
with published experimental results is obtained, without the need of 
introducing fitting parameters.
\end{abstract}
\maketitle 
In the recent literature, the topic of shot noise suppression
in mesoscopic structures
has received significant attention, as a result of the formulation of several
theoretical predictions \cite{onethird,beejala,coulbee,agam,
silvestrov} and the subsequent experimental confirmation
\cite{onethex,ober1} of some of such predictions. All of the shot noise 
suppression phenomena
are the result of correlations between charge carriers that decrease the
variance of the random process corresponding to the elementary charges
crossing the device: such correlations may result either from Fermi exclusion
or from Coulomb interaction, and for their investigation powerful theoretical
methods have been developed, ranging from Random Matrix Theory \cite{beerev}
to semiclassical techniques~\cite{blantsuk}. The convergence, in terms of 
shot noise suppression, between the 
results of quantum and semiclassical approaches has been explained, using a 
voltage probe technique, by van Langen and B\"uttiker~\cite{langen}, who have 
shown that dephasing phenomena do not have effects on the noise power spectral
density. This result appears to be valid also in the presence of non DC bias
\cite{polianski}. 
A detailed discussion of the 
fundamentals of shot noise in mesoscopic conductors can be found in 
ref.~\cite{blabuet}.

Particular interest has been raised by the 
so-called ``chaotic cavities'', mesoscopic regions delimited
by input and output constrictions that are much smaller than the cavities
themselves. 
Jalabert {\it et al.}~\cite{beejala} showed that in the case of symmetric 
apertures
noise is suppressed down to 1/4 of its full shot value: this theory received
experimental confirmation \cite{ober1} in 2001, and, more recently, the
noise behavior of a chaotic cavity in an orthogonal magnetic field has been
measured \cite{ober2}, observing a somewhat linear reduction of the Fano 
factor as the
magnetic field is increased. In ref.~\cite{ober2} the authors motivate
the decrease of the Fano factor as the magnetic field is increased with the
reduction of the portion of the cavity area explored by the electrons, and 
rely on a fitting parameter, the quantum scattering time, which is also used
to explain the behavior of the Fano factor with no magnetic field as the 
apertures are made wider.
They also attribute the differences between the variation of the conductance 
and that of the Fano factor to a transition between ``quantum'' and 
``classical'' chaos. 

We provide an alternative, intuitive interpretation of the behavior in the
presence of a magnetic field, based on the comparison 
between the classical cyclotron diameter and the constriction width, instead
of the cavity dimensions, which 
is supported by the results of the numerical simulations that we present in 
the following. Furthermore, we argue that there is no significant 
contribution of ``classical'' chaos to the observed phenomena, in particular
we suggest a different explanation of the preserved additivity of the 
constriction resistances up to 40 propagating modes,
in the case of no magnetic field. Such an explanation is based on assuming
some degree of nonideality in the experiment, specifically partial 
thermalization to the lattice temperature of the electron gas in the cavity.

\begin{figure}
\includegraphics[clip,angle=0,width=8.4cm]{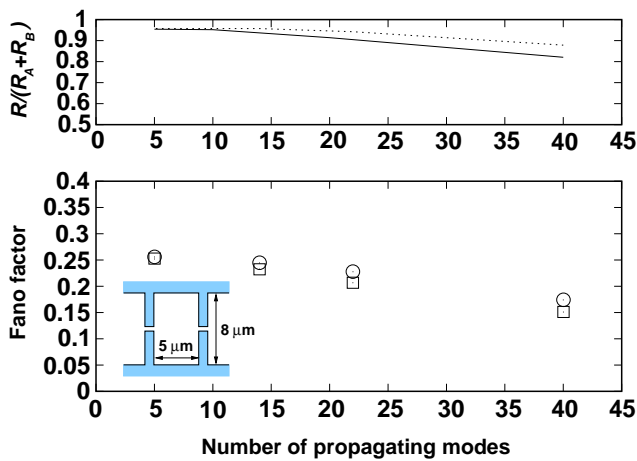}
\caption{Ratio of the cavity resistance to the sum of the constriction 
resistances {\it vs.} the number of modes propagating in the constrictions
(upper panel): the solid line is for our rectangular cavity and the dotted 
line for a stadium-shaped cavity with a central constant-width region 
1~$\mu$m long. Numerical (empty circles) and experimental (empty squares, from
Oberholzer {\it et al.}~\cite{ober2}) results for the Fano factor as a 
function of the 
number of modes propagating in the constrictions, for no applied magnetic 
field (lower panel). The inset contains a sketch of the model cavity.}
\label{b0}
\end{figure}

In most of our calculations, we consider model cavities defined by 
hard walls and with a rectangular shape (see inset of fig.~\ref{b0}), in order 
to keep the computational
time within reasonable limits, running a few checks for structures with
different geometries: only in few particular cases 
small differences were observed, which will be detailed in the discussion of
the results. 
We wish to point out that a classically chaotic shape is not needed to 
achieve the known results for shot noise suppression. This conclusion cannot 
be found in unequivocal and explicit form in the existing literature, except
for the very recent paper by Aigner {\it et al.}\cite{aigner}, who have shown 
that
the distribution of transmission eigenvalues (and therefore the Fano factor)
in a cavity with narrow constrictions does not depend appreciably on
its shape.
 
However, the theoretical basis needed to reach such a conclusion had already 
been developed by several authors.
For example, we can apply to a single cavity, in the absence of a magnetic
field, the analytic procedure 
discussed by Oberholzer {\it et al.}~\cite{obermulti} for a series of 
cavities. The cavity is supposed
to act as an elastic quasi-reservoir, effectively decoupling the two 
constrictions:
by balancing the incoming and outgoing 
electron fluxes at each energy~\cite{obermulti}, one obtains, in the 
hypothesis of symmetric and narrow constrictions, the well-known result of 
1/4 for the Fano factor. 

We wish to point out that to arrive at this result, no hypothesis of 
a classically chaotic shape is needed: the discontinuities represented
by the transitions between the cavity and the leads produce the mode mixing 
needed to generate a uniform occupancy of all the states at the same energy
({\it i.e.} the property defining an elastic quasi-reservoir),
which, along with symmetry and integer transmission of the constrictions,
represents the essential condition needed to achieve the 1/4 suppression
factor~\cite{obermulti}.

This is the reason why we prefer to 
define the subject of our investigation as ``mesoscopic cavities,'' rather than
``chaotic cavities,'' in order to avoid possible confusion with classical 
chaotic dynamics. Quantum diffraction occurring only at the input and output 
openings leads to multiple trajectories that create the equivalent of a 
chaotic behavior. 

Our numerical technique is based on the evaluation of the transmission matrix 
of the structure being considered with the scattering matrix approach, which 
is relatively easy to implement and exhibits good numerical stability, also in 
the case of high values of the magnetic field $B$. 

In particular, we consider a 
2-dimensional Schr\"odinger equation in the $x$-$y$ plane, with $x$ being
the direction of electron propagation (and thus $z$ the direction of the 
uniform orthogonal magnetic field). We choose a vector potential with a
single nonzero component along the longitudinal direction $x$ (with a
value $-By$) and we subdivide the structure into a number of 
transverse slices, in each of which the scalar (confinement) potential
can be assumed to be independent of $x$. In each slice we expand the
transverse eigenfunctions onto a basis made up of the transverse 
eigenfunctions for $B=0$; the coefficients of this expansion and the associated
longitudinal wave vectors are found solving an eigenvalue problem 
\cite{tamuraando}. Then, with a mode-matching technique, we 
compute the scattering matrices of the sections extending from 
the middle of each slice to the middle of the following slice (and thus 
containing only one discontinuity of the potential). 
Composing the scattering matrices of all the sections~\cite{datta}, we find the
S-matrix (and, as a submatrix, the transmission matrix $t$ with elements 
$t_{nm}$) of the overall
structure.

The value of the conductance and of the shot noise power spectral
density can be then computed by means of the relations~\cite{butprl}
\begin{equation}
G=\frac{2\,e^2}{h}\,\sum_{n,m}\left|t_{nm}\right|^2 =
\frac{2\,e^2}{h}\,\sum_j w_j 
\end{equation}
and
\begin{equation}
S_I=4\,\frac{e^3}{h}\,|V|\,\sum_j w_j\,(1-w_j)\ ,
\label{uu} 
\end{equation}
where the $w_j$'s are the eigenvalues of the matrix ${t\,t^{\dag}}$, $e$ is 
the electron charge, $h$ is Planck's constant, and $V$ is
the externally applied voltage. Since the power spectral density of full
shot noise is given by $S_{I}^{fs}=2 e |I|$ ($I$ being the average current 
through
the device) the Fano factor $\gamma$ can be written 
as
\begin{equation}
\gamma={\frac {\sum_j w_j(1-w_j)} {\sum_j w_j}}\, .
\label{fanoeq}
\end{equation}
For a small number of modes propagating through the constrictions, quantum 
interference effects lead to wide relative fluctuations, as a function of 
energy, of both the numerator and the denominator of this expression. 
Averaging is therefore needed, and 
care must be taken to perform it correctly: in order to measure shot noise,
the applied voltage $V$ must be much larger than $kT/e$ ($k$ being the 
Boltzmann constant and $T$ the absolute temperature). Based on the detailed
expression of the shot noise power spectral density at finite temperature
provided by B\"uttiker~\cite{buttipap}, in the case of $eV \gg kT$ a good 
approximation of the shot noise term is given by a uniform average over
an interval $eV$ around the Fermi level of eq.~(\ref{uu}). An analogous 
conclusion, {\it i.e.}, uniform averaging over the interval $eV$, can be 
drawn for the conductance term. We remark that
it is essential that averaging be performed separately on the numerator and 
the denominator before taking the ratio (as in the actual measurement 
procedure).  

Each of our data points has 
been obtained by averaging over the noise power spectral density and the 
conductance for 41 different energy 
values in an interval of width 0.22~meV around the Fermi energy.
In all of the following calculations we consider a Fermi energy $E_f$
of 9.134~meV and the effective mass of
gallium arsenide, $m^*=0.067 m_0$, with $m_0$ being the free electron mass 
(values consistent with the experiments by Oberholzer {\it et al.}). 

\begin{figure}
\includegraphics[clip,angle=0,width=8.4cm]{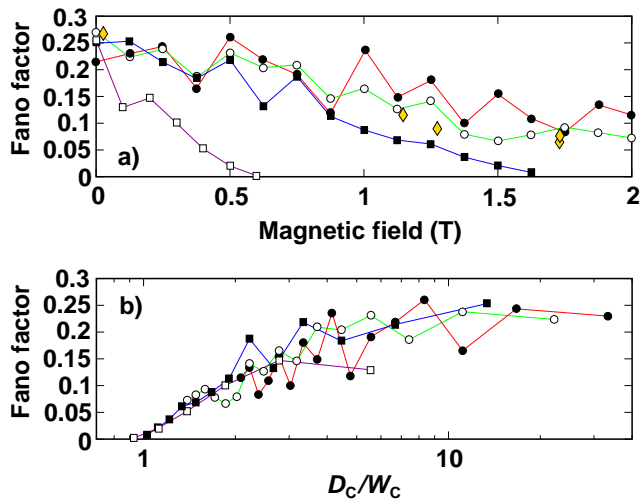}
\caption{(a) Fano factor as a function of the magnetic field computed for a
mesoscopic cavity with constrictions of 40~nm (solid circles), 
60~nm (empty circles), 100~nm (solid
squares), 300~nm (empty squares). The diamonds represent the experimental
data by Oberholzer {\it et al.}; (b) same data as in (a), but plotted as
a function of the ratio of the classical cyclotron diameter $D_C$ to the
constriction width $W_C$. }
\label{magnetoplot}
\end{figure}

This numerical approach has been used for the simulation of shot noise
suppression in a rectangular cavity,
first in the absence of a magnetic field. In particular, we show
that the expected shot noise suppression is achieved with a simple
hard-wall model defining a rectangular cavity, 
5~$\mu$m long and
8~$\mu$m wide (corresponding to the lithographic dimensions of the cavity in
refs.~\cite{ober1,ober2}), with symmetric constrictions whose width is
chosen on the basis of the desired number of propagating modes.

As constrictions are made wider (reaching a condition in which the
analytical theory yielding the value 1/4 for the suppression factor
is not applicable any longer), we observe a decrease of the Fano factor,
until it drops to zero as their width equals the cavity width
and the structure becomes a noiseless perfect quantum wire. In the 
lower panel of fig.~\ref{b0},
we report the Fano factor from our numerical calculations (empty circles) as a
function of the number of propagating modes: 
empty squares represent the
experimental data obtained by Oberholzer {\it et al.} \cite{ober2}, and the
agreement appears to be good, without the need for any fitting parameter.
In the upper panel of fig.~\ref{b0} we report the ratio of the overall
cavity resistance to the sum of the constriction resistances, as a function of
the number of propagating modes with a solid line for our rectangular cavity
and with a dotted line for a stadium-shaped cavity (8 $\mu$m wide and 
9 $\mu$m long, since a horizontal stadium cannot be made with the same aspect 
ratio as that of the experimental cavity): such a quantity drops below unity, 
in disagreement with the experiment.
We show also the calculation for the stadium-shaped cavity in order 
to exclude the possibility of a significant role played by a classically 
chaotic shape. 
Also the inclusion of disorder in the cavity (data for which are not 
reported here, due to space constraints) would not provide a valid 
justification of resistance behavior, since it would raise both resistance and 
noise.

We suggest, as a possible explanation, that such an additivity over the whole 
measurement range can be the
result of partial electron thermalization: as constrictions 
are made wider and electron diffraction is reduced, the total resistance
of the cavity should start decreasing below the sum of the individual 
constrictions, according to the outcome of our ballistic calculation, while 
it would remain constant, independent of constriction width, if the cavity
were a real reservoir, with electrons in thermal equilibrium with the lattice.
If full thermalization of the electrons did occur, shot noise would also 
disappear (as long as the constrictions have integer transmission); a partial
thermalization would instead determine an increase of the resistance toward
the classical sum rule and, at the same time, a decrease of the noise, with
respect to the ballistic result: this is consistent with the comparison 
between our results and those from the experiment.

We have then moved on to the 
simulation of the behavior of a cavity in the presence of a magnetic field
orthogonal to the plane of the device, with a value
up to a few tesla.

\begin{figure}
\includegraphics[clip,angle=0,width=8.4cm]{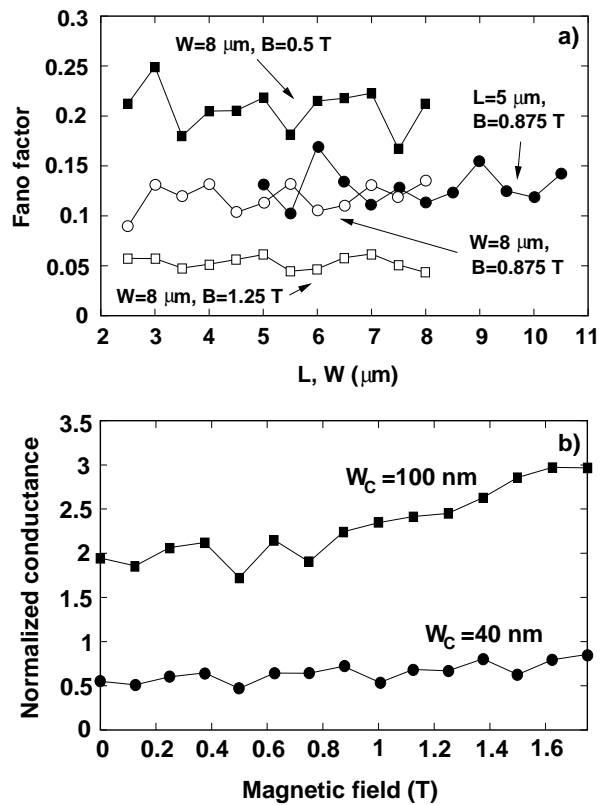}
\caption{(a) Fano factor as a function of the cavity width $W$ (solid circles) 
and length $L$ (remaining plots) for 100~nm symmetric constrictions;
(b) conductance through the cavity, normalized with respect
to the conductance quantum, as a function of the orthogonal magnetic field,
for two values of the constriction widths.}
\label{figwl}
\end{figure}

Results for the Fano factor are reported in fig.~\ref{magnetoplot}(a)
for symmetric constrictions with a width of 40~nm (solid circles), 60~nm (empty
circles), 100~nm (solid squares), and 300~nm (empty squares). The Fano factor
exhibits a decrease as a function of the
magnetic field, with a rate depending on the width of the constrictions
and, therefore, on the number of modes propagating through such constrictions.
The smaller the number of propagating modes the larger the fluctuations,
due to the more significant relative contribution from quantum interference:
for the narrowest constrictions residual fluctuations survive our averaging 
procedure.

The only available experimental data, from ref.~\cite{ober2}, are for a single 
mode propagating
through the constrictions \cite{Obercomm}, which corresponds, in our model,
to a constriction width of 40 nm. Experimental results from \cite{ober2},
represented with diamonds, lie slightly below the theoretical data for 
the 40~nm constrictions 
and close to those for the 60~nm constrictions. Thus
our simulations tend to slightly overestimate the Fano factor, for narrow
constrictions and large $B$ values. This can be attributed,
in particular for a single propagating mode, to the specific scattering 
properties of hard-wall constrictions, which
tend to create a set of modes inside the cavity, some of which are almost
completely
reflected at the exit constriction, contrary to gradual adiabatic constrictions
(as we have verified with numerical 
simulations on a structure with openings defined as in ref.~\cite{andoadiab}).

In fig.~\ref{magnetoplot}(b) we report the same data as that of 
fig.~\ref{magnetoplot}(a), plotted as a function of the ratio of the 
classical
cyclotron diameter $D_C$ (twice the cyclotron radius) to the constriction 
width $W_C$: it is apparent
that results for different constriction widths are essentially superimposed,
thus demonstrating our conjecture that $D_C/W_C$ is the actually relevant 
parameter 
defining the level of shot noise suppression, simply because it determines
the amount of scattering, and therefore diffraction, that occurs at the
constrictions.  Indeed, also the r.h.s. of 
eq.~(4) of ref.~\cite{ober2} is proportional to $D_C$ and inversely 
proportional to $W_C$: however it would also be proportional to the 
perimeter (or length), a dependence which is not observed in our results.



Running our simulations for different values of the cavity width and 
length (see fig.~\ref{figwl}(a)), we have
noticed that, in the regime of narrow constrictions and noise suppression 
due to the magnetic field, the Fano factor has only a mild and nonmonotonic 
dependence on such parameters (unless the constrictions are
rather wide, and we are not in the regime of interest any longer), arguably 
resulting just from fluctuations due to interference effects.
This is not in agreement with the interpretation of the phenomenon
proposed by Oberholzer {\it et al.}~\cite{ober2}. Starting from a
first-order expression involving the ratio of a quantum scattering
time to the dwell time in the cavity, they suggest that the Fano
factor should depend on the cavity area available for electron
motion and they explain the shot noise suppression with a
reduction of this area as a consequence of the formation of
cyclotron orbits. 

From the results of our numerical calculations, we propose a different and
intuitive interpretation, directly based on the comparison of the cyclotron
diameter with constriction width. In a semiclassical picture with skipping 
orbits crawling along the cavity walls, if the cyclotron diameter $D_C$ is much
larger than the constriction width $W_C$, an electron impinging against such
a constriction is likely to be reflected, thereby leading to diffraction.
On the contrary, if $D_C$ is smaller than $W_C$, it is very likely that the
electron traverses the two constrictions without undergoing reflections, so
that shot noise is strongly suppressed (in the limit of no scattering, the 
Fano factor would drop to zero).


We have also computed the conductance of the cavity as a function of 
magnetic field, observing the transition, measured in ref.~\cite{ober2},
between a condition, for $B=0$, in which it equals one half of the conductance
of each constriction, and that for a large magnetic field, in which edge
states are formed and it reaches the value of a single constriction.
Results are reported in fig.~\ref{figwl}(b), where the conductance for 
a constriction width of 40~nm (1 propagating mode) starts at 0.5 units 
and increases toward 1 unit as in the experimental results of 
ref.~\cite{ober2}, while
the curve for $W_C=100$~nm (4 propagating modes for $B=0$) has an initial
value of 2 and a value of 3 for a large magnetic field.
This latter value is not 4, as expected, because the total number of 
modes propagating through the constrictions decreases down to 3 at about 0.5~T.
We notice that the experimental data for conductance are in this case more 
closely reproduced by our ballistic simulation, as can be explained by 
the fact that in the presence of a magnetic field the effective area of the 
cavity and the time spent by each electron inside it are reduced, 
and therefore thermalization is decreased. We also wish to point out
that, for the explanation of the observed phenomena, there is no need to 
invoke the presence of irregularities or defects in the cavity: indeed, as
in the case of no magnetic field, they
would lead to an increase of noise, due to the resulting electron diffraction. 
   
In conclusion, we have presented a numerical simulation of shot noise 
suppression in mesoscopic cavities, reproducing the results of recent 
experiments without the need for fitting parameters. In particular, we 
get very good agreement with the measured behavior of the Fano factor as
a function of the width of the constrictions and, although with fluctuations
due to interference effects,  we are able to closely reproduce the dependence 
of the Fano factor on magnetic field. From our results, we observe that the 
reduction of the Fano factor with increasing magnetic field can be simply
explained as the consequence of the decrease in the ratio of the cyclotron 
diameter to the constriction width, which leads to progressively improved 
transmission through the apertures, and therefore suppression of diffraction.
 
This work has been supported by the Italian 
Ministry for Education, University and Research, through the FIRB project
``Nanotechnologies and nanodevices for the information
society'' and through the PRIN ``Excess noise in nanoscale
devices.''


\begin{thebibliography}{999}

\bibitem{onethird} Beenakker~C.~W.~J. and B\"uttiker~M., {\it Phys.~Rev.~B}, 
{\bf 46}
(1992) 1889.

\bibitem{beejala} Jalabert~R.~A., Pichard~J.~L. and Beenakker~C.~W.~J.,
{\it Europhys.~Lett.}, {\bf 27} (1994) 255.


\bibitem{coulbee} Beenakker~C.~W.~J., {\it Phys.~Rev.~Lett.}, 
{\bf 82} (1999) 2761.

\bibitem{agam} Agam~O., Aleiner~I. and Larkin~A., {\it Phys.~Rev.~Lett.}, 
{\bf 85}
 (2000) 3153.

\bibitem{silvestrov} Silvestrov~P.~G., Goorden~M.~C. and Beenakker~C.~W.~J.,
{\it Phys.~Rev.~B}, {\bf 67} (2003) 241301(R).

\bibitem{onethex} Henny~M., Oberholzer~S., Strunk~C. and Sch\"onenberger~C.,
{\it Phys.~Rev.~B}, {\bf 59} (1999) 2871.

\bibitem{ober1} Oberholzer~S., Sukhorukov~E.~V., Strunk~C.,
Sch\"onenberger~C., Heinzel~T. and Holland~M.,
{\it Phys.~Rev.~Lett.}, {\bf 86} (2001) 2114.

\bibitem{beerev} Beenakker~C.~W.~J., {\it Rev.~Mod.~Phys.}, 
{\bf 69} (1997) 731.

\bibitem{blantsuk}
Blanter~Ya.~M. and Sukhorukov~E.~V., {\it Phys.~Rev.~Lett.}, 
{\bf 84} (2000) 1280.

\bibitem{langen} van~Langen~S.~A. and B\"uttiker~M., 
{\it Phys.~Rev.~B}, {\bf 56} (1997) R1680.

\bibitem{polianski} Polianski~M.~L., Samuelsson~P. and B\"uttiker~M.,
{\it Phys.~Rev.~B}, {\bf 72} (2005) 161302(R).

\bibitem{blabuet} Blanter~Ya.~M., B\"uttiker~M., {\it Phys.~Rep.}, {\bf 336}
(2000) 1.

\bibitem{ober2} Oberholzer~S., Sukhorukov~E.~V. and Sch\"onenberger~C.,
{\it Nature}, {\bf 415} (2002) 765.

\bibitem{aigner} Aigner~F., Rotter~S. and Burgd\"orfer~J.,
{\it Phys.~Rev.~Lett.}, {\bf 94} (2005) 216801.



\bibitem{obermulti} Oberholzer~S., Sukhorukov~E.~V., Strunk~C. and
Schonenberger~C., {\it Phys.~Rev.~B.}, {\bf 66} (2002) 233304.

\bibitem{tamuraando} Tamura~H. and Ando~T., {\it Phys.~Rev.~B}, 
{\bf 44} (1991) 1792.

\bibitem{datta} Datta~S., {\it Electronic Transport in Mesoscopic Systems}
(Cambridge University Press, Cambridge) 1995, p.~125.
 
\bibitem{butprl} B\"uttiker~M., {\it Phys.~Rev.~Lett.}, {\bf 65} (1990) 2901.

\bibitem{buttipap} B\"uttiker~M., {\it Phys.~Rev.~B}, {\bf 46} (1992) 12485.

\bibitem{Obercomm} Oberholzer~S., private communication

\bibitem{andoadiab} Ando~T., {\it Phys.~Rev.~B}, {\bf 44} (1991) 8017.




\end{thebibliography}
\end{document}